# Weak value picture on quantum observables: gauge-invariant vector potentials

Sunkyu Yu[1,2,†], Xianji Piao[1] and Namkyoo Park[1*]

[1]*Photonic Systems Laboratory, Department of Electrical and Computer Engineering, Seoul National University, Seoul 08826, Korea*
[2]*Intelligent Wave Systems Laboratory, Department of Electrical and Computer Engineering, Seoul National University, Seoul 08826, Korea*

E-mail: [†]sunkyu.yu@snu.ac.kr, [*]nkpark@snu.ac.kr

**Abstract** The conservation of physical quantities under coordinate transformations, known as gauge invariance, has been the foundation of theoretical frameworks in both quantum and classical theory. The finding of gauge-invariant quantities has enabled the geometric and topological interpretations of quantum phenomena with the Berry phase, or the separation of quark and gluon contributions in quantum chromodynamics. Here, with an example of quantum geometric quantities—Berry connection, phase, and curvature—we extract a new gauge-invariant quantity by applying a "weak value picture". By employing different pre- and post-selections in the derivation of the Berry phase in the context of weak values, we derive the gauge-invariant vector potential from the Berry connection that is originally gauge-dependent, and show that the obtained vector potential corresponds to the weak value of the projected momentum operator. The local nature of this quantity is demonstrated with an example of the Aharonov-Bohm effect, proving that this gauge-invariant vector potential can be interpreted as the only source of the Berry curvature in the magnetic field. This weak value decomposition approach will lead to the extraction of new measurable quantities from traditionally unobservable quantities.

## 1. Introduction

Gauge invariance, having deep roots in the foundation of electromagnetism [1] since H. Weyl's seminal paper [2,3], constitutes one of the pillars of modern physics. The finding of invariant quantities under gauge transformation allows for the composition of an observable system with measurable quantities [4]. The most introductory and historical example is found in classical electromagnetics [5], as shown in the comparison between the gauge-invariant, measurable quantities of the fields (**E**, **H**) and the gauge-dependent, non-unique quantities of the potentials (**A**, *V*). Other examples include the separation of quark and gluon contributions to the nucleon spin, using gauge-invariant decomposition in quantum chromodynamics [6].

One of the most important examples in finding gauge-invariant quantities is shown in the discovery of the Berry phase [7,8], which describes geometrically defined phase accumulation based on the adiabatic change of a system in a specific parameter space. In his original paper [7] that provided a framework for the "geometrization" of physical problems [8,9] and the extension to topological theory [10,11], M. V. Berry focused on the vector-potential term known as the Berry connection, $\mathbf{A}_n = i\langle u_n|\nabla_\mathbf{R}|u_n\rangle$, where $|u_n\rangle$ is the $n^\text{th}$ system eigenstate, **R** is the parameter space, and $|u_n\rangle$ and the gradient operator $\nabla_\mathbf{R}$ are well defined in the complex Hilbert space. Similar to the electromagnetic vector potential **A**, Berry connection $\mathbf{A}_n$ is gauge-dependent for the transformation of the wavefunction, as $\mathbf{A}_n \rightarrow \mathbf{A}_n - \nabla_\mathbf{R}\xi_n(\mathbf{R})$ with $|u_n\rangle \rightarrow exp[i\xi_n(\mathbf{R})]|u_n\rangle$. Therefore, the Berry connection had been ignored for a long time prior to the identification of the Berry phase [7,8] – the *cyclic integral* of $\mathbf{A}_n$ in the **R**-space associated with gauge invariance. Although the concept of the Berry phase has been extended to non-adiabatic (Aharonov-Anandan phase) [12] and non-cyclic (Pancharatnam phase) [13-15] evolution in the ray space, the Berry connection $\mathbf{A}_n$ or other generalized geometric connections [12-14], which are the sources of quantum geometric quantities including Berry phase and curvature, have remained gauge-dependent, similar to **A** in electromagnetics.

To suggest a new viewpoint on handling gauge invariance and therefore defining quantum observables, consider the traditional representation of the Berry connection $\mathbf{A}_n = \langle u_n | i\nabla_{\mathbf{R}} | u_n \rangle$; $\mathbf{A}_n$ is the expectation value of $|u_n\rangle$ for the parameter-space "momentum-like" operator $i\nabla_{\mathbf{R}}$. The Berry connection thus follows the mathematical form of the quantum measurement for $i\nabla_{\mathbf{R}}$ with the same pre- ($|u_n\rangle$) and post-selected ($\langle u_n|$) states, though the operator $i\nabla_{\mathbf{R}}$ is gauge-dependent and therefore is not an observable as we discuss later. This perspective inspires the *revisiting* of $\mathbf{A}_n$ with the concept of a "weak value" having different pre- and post-selected states [16-18], which describes the information of a physical quantity through a weak measurement. Because such a weak measurement with a weakly coupled system and probe permits an almost undisturbed initial state [17], we can envision the measurement of a certain expectation value through a set of weak values from multiple weak measurements. In terms of the measurements related to gauge invariance, the question then arises: Does each "partial" weak value have the same gauge condition as that of the original expectation value? The answer is directly connected to the extraction of new measurable quantities from an unobservable one in terms of the concept of a weak value. Notably, although the weak value of the gauge-invariant geometric *phase* has been studied [19-21], the application of weak values to the gauge-dependent, unobservable Berry *connection* has not been considered.

In this paper, as an example of finding quantum observables using the concept of weak values, we extract the gauge-invariant vector potential from the weak value representation of the Berry connection. By employing different pre- and post-selected states [16-18], we revisit the derivation of the Berry phase, which yields the decomposition of the Berry connection $\mathbf{A}_n$ into gauge-dependent and gauge-invariant weak-value vector potentials. This decomposition provides a measurable condition for the mutually projected momentum weak value in the parameter space $\mathbf{R}$, for an arbitrary definition of the post selection (or, an arbitrary weak measurement condition). As a representative example, we apply this approach to the Aharonov-Bohm effect to illustrate the local nature of the newly found gauge-invariant weak value. Finally, we discuss the topological nature of the gauge-invariant weak value, showing that this gauge-invariant weak value becomes the only source of the gauge-invariant Berry curvature in the Aharonov-Bohm effect with a *nonzero* magnetic field, or more generally, in the post-selection with parameter state $\langle \mathbf{R}|$. By revealing a new measurable geometric quantity in the form of weak value vector potentials, our results provide novel insight into the measurements of geometric and topological properties in quantum mechanics [8,10] and optics [9,11,22].

## 2. Weak-value picture on decomposing Berry connection

Following the derivation of the Berry phase [7,8,23], consider a time-dependent Hamiltonian $H = H(t)$. We focus on the Hamiltonians which can be described by a set of parameters $\mathbf{R} = (R_1, R_2, \ldots)$, as $H = H(\mathbf{R}(t))$. We also restrict our discussion to the complex Hilbert space $\mathbf{R}$, such as the Euclidean space with position states or its reciprocal space with momentum states. The state of the Hilbert space can then be represented by the state ket $|\mathbf{R}\rangle$. The Hamiltonian for a spinless particle under a uniform magnetic field is a representative example of such Hamiltonians: the spatially varying form of the Hamiltonian due to the Landau gauge [23,24]. The Hamiltonians for a broad class of topological phenomena [25,26], the momentum-dependent Hamiltonians [27], and wave behaviors in non-Hermitian [28,29] or disordered [30,31] systems can also be described by the parameter space $\mathbf{R}$.

Using this set of parameters $\mathbf{R}$, the time-dependent evolution of the Hamiltonian $H = H(\mathbf{R}(t))$ is described by the time-dependent parameters $\mathbf{R}(t) = (R_1(t), R_2(t), \ldots)$. For adiabaticity, we consider non-degenerate eigenstates $|u_n(\mathbf{R}(t))\rangle$ of the Hamiltonian $H(t)$. Also, for simplicity, we assume the normalized eigenstate in time $\langle u_n(t)|u_n(t)\rangle = 1$. When a state $|\psi(\mathbf{R}(t))\rangle$ is initially in the $n^{\text{th}}$ eigenstate $|u_n(\mathbf{R}(t))\rangle$ satisfying $H(\mathbf{R})|u_n(\mathbf{R})\rangle = E_n(\mathbf{R})|u_n(\mathbf{R})\rangle$, the exact time-varying solution of $|\psi(\mathbf{R})\rangle$ becomes [23]

$$|\psi(\mathbf{R})\rangle = e^{i\theta_n(t)} e^{i\gamma_n(t)} |u_n(\mathbf{R})\rangle + \varepsilon \sum_{m \neq n} c_m(t) |u_m(\mathbf{R})\rangle, \qquad (1)$$

where $\theta_n(t) = -(1/\hbar)\int_0^t E_n(\mathbf{R}(t'))dt'$ is the dynamical phase, $\gamma_n(t)$ is the Berry phase, $\varepsilon$ denotes the deviation from the adiabatic limit, and $c_m$ is the transition coefficient to the $m^{\text{th}}$ eigenstate. With the time-dependent Schrödinger equation $i\hbar\partial_t|\psi(\mathbf{R})\rangle = H(\mathbf{R})|\psi(\mathbf{R})\rangle$, it follows that

$$\left(\frac{\partial}{\partial t} + i\frac{\partial\gamma_n(t)}{\partial t}\right)|u_n(\mathbf{R})\rangle = -e^{-i(\theta_n+\gamma_n)}\varepsilon\sum_{m\neq n}\left[\left(\frac{i}{\hbar}c_m E_m + \frac{\partial c_m}{\partial t}\right)|u_m(\mathbf{R})\rangle + c_m\frac{\partial}{\partial t}|u_m(\mathbf{R})\rangle\right]. \qquad (2)$$

The second-order term of $c_m\partial_t|u_m\rangle$ can be ignored with the adiabatic condition [23]. In the derivation of the Berry phase, the multiplication by the initial bra eigenstate $\langle u_n(\mathbf{R})|$ in Eq. (2), which corresponds to the identical pre-selection and post-selection in the weak value definition [16-18,32], leads to $\partial_t\gamma_n(t) = i\langle u_n|\partial_t|u_n\rangle$. Note that the time-varying state $|\psi(\mathbf{R})\rangle$ in Eq. (1) does not have to be normalized because all of the coefficients $c_m$ ($m \neq n$) vanishes when the bra eigenstate $\langle u_n(\mathbf{R})|$ is applied to Eq. (2).

Applying the chain rule $\partial_t = (\partial \mathbf{R}/\partial t)\nabla_\mathbf{R}$ and integrating the result $\partial_t \gamma_n(t) = i\langle u_n|\partial_t|u_n\rangle$ with respect to time then provides the Berry phase of $\gamma_n(C) = \oint \mathbf{A}_n d\mathbf{R}$, which is gauge-invariant modulo $2\pi$ for the cyclic integral along the closed loop C, where $\mathbf{A}_n = i\langle u_n|\nabla_\mathbf{R}|u_n\rangle$ is the Berry connection [7,8,23].

To interpret the Berry connection $\mathbf{A}_n$ as the expectation value of the transient state $|u_n\rangle$, we first need to develop the notion of the canonical momentum operator in the complex-Hilbert "parameter space $\mathbf{R}$" from the definition of $\mathbf{R}$. For the wavefunction $|\psi(\mathbf{R})\rangle$ defined in the $N$-dimensional parameter space $\mathbf{R} = (R_1, R_2, ..., R_N)$, we introduce the $\mathbf{R}$-translation operator $\mathbf{T_R}(\Delta\mathbf{R})$, which leads to the transformation of a wavefunction for the changes in parameters $\Delta\mathbf{R} = (\Delta R_1, \Delta R_2, ..., \Delta R_N)$, as $\mathbf{T_R}(\Delta\mathbf{R})|\psi(\mathbf{R})\rangle = |\psi(\mathbf{R} - \Delta\mathbf{R})\rangle$, such as spatial translation $\mathbf{T_x}(\Delta\mathbf{x})|\psi(\mathbf{x})\rangle = |\psi(\mathbf{x} - \Delta\mathbf{x})\rangle$ for $\mathbf{R} = \mathbf{x}$, time translation $\mathbf{T}_t(\Delta t)|\psi(t)\rangle = |\psi(t - \Delta t)\rangle$ for scalar $R = t$, and momentum translation $\mathbf{T_k}(\Delta\mathbf{k})|\psi(\mathbf{k})\rangle = |\psi(\mathbf{k} - \Delta\mathbf{k})\rangle$.

Following the definition of the canonical momentum operator as the generator of spatial translation, we define the canonical momentum operator in the $\mathbf{R}$-space $\mathbf{p_R}$ as the "generator of parameter translation", where $p_\mathbf{R}^m$ is the momentum operator for the "$m^{\text{th}}$ parameter" $R_m$:

$$p_\mathbf{R}^m = i\hbar \lim_{\Delta R_m \to 0} \frac{\mathbf{T_R}(\Delta R_m) - \mathbf{I}}{\Delta R_m}, \quad (3)$$

with the identity operator $\mathbf{I}$. The application of the $m^{\text{th}}$ momentum operator to the wavefunction then gives $p_\mathbf{R}^m|\psi(\mathbf{R})\rangle = -i\hbar(\partial/\partial R_m)|\psi(\mathbf{R})\rangle$, and thus, the canonical momentum operator in the $\mathbf{R}$-space is defined as $\mathbf{p_R} = -i\hbar\nabla_\mathbf{R}$. Because the *unitary* $\mathbf{R}$-translation operator $\mathbf{T_R}(\Delta\mathbf{R})$ guarantees the Hermitian operator $\mathbf{p_R}$ according to Stone's theorem (or, see Appendix A for the simple verification), the application of the weak value theory based on the Hermitian operator [16-18,32] is thus allowed for the unitary translational operator $\mathbf{T_R}(\Delta\mathbf{R})$ in the $\mathbf{R}$-space, covering a wide range of fundamental parameter translations, such as the spatial translation ($\mathbf{R} = \mathbf{x}$) for the real-space momentum and the time translation ($\mathbf{R} = t$) for the Hamiltonian.

For the weak-value formulation using the Hermitian operator $\mathbf{p_R}$, we introduce a new degree of freedom: a *generalized* and *arbitrary* post-selection with $\langle\varphi(\mathbf{R})|$ to the Berry connection, instead of the conventional post-selection with $\langle u_n|$. The time derivative of the Berry phase $\partial_t\gamma_n(t)$ can be re-expressed as $\partial_t\gamma_n = i\langle\varphi|u_n\rangle\langle u_n|\partial_t|u_n\rangle/\langle\varphi|u_n\rangle$ with the condition of $\langle\varphi|u_n\rangle \neq 0$, which derives the following form from the identity operator $\mathbf{I} = \sum|u_m\rangle\langle u_m|$, as (see Appendix B for the direct application of the post-selection to the state of the time-dependent Schrödinger equation):

$$\frac{\partial \gamma_n(t)}{\partial t} = \frac{i\langle\varphi|\frac{\partial}{\partial t}|u_n\rangle - i\sum_{m \neq n}\langle\varphi|u_m\rangle\langle u_m|\frac{\partial}{\partial t}|u_n\rangle}{\langle\varphi|u_n\rangle}. \quad (4)$$

By applying the chain rule of $\partial_t = (\partial\mathbf{R}/\partial t)\nabla_\mathbf{R}$ to Eq. (4) and integrating the result with respect to time, we arrive at the "weak value picture" of the Berry connection $\mathbf{A}_n$ for $\gamma_n(C) = \oint \mathbf{A}_n d\mathbf{R}$, as the sum of two weak-value vector potentials $\mathbf{A}_n = \mathbf{A}_n^S + \mathbf{A}_n^{MP}$, which is the first main result of this paper:

$$\mathbf{A}_n^S = i\frac{\langle\varphi|\nabla_\mathbf{R}|u_n\rangle}{\langle\varphi|u_n\rangle}, \quad (5)$$

$$\mathbf{A}_n^{MP} = -i\sum_{m \neq n}\frac{\langle\varphi|u_m\rangle\langle u_m|\nabla_\mathbf{R}|u_n\rangle}{\langle\varphi|u_n\rangle} = \sum_{m \neq n}\mathbf{A}_n^{MP\text{-}m}. \quad (6)$$

We note that this weak value picture provides an alternative understanding of the Berry connection $\mathbf{A}_n$ as the sum of weak values for the operators $\nabla_\mathbf{R}$ and $|u_m\rangle\langle u_m|\nabla_\mathbf{R}$ ($m \neq n$).

Although $\mathbf{A}_n^S$, $\mathbf{A}_n^{MP}$, and $\partial_t\gamma_n$ in Eqs. (4-6) are defined for every time $t$, we focus on the closed loops in the parameter space to examine the Berry phase, the gauge-invariant quantity. Using the relationship $\mathbf{A}_n = \mathbf{A}_n^S + \mathbf{A}_n^{MP}$, the Berry phase defined along the closed loop C can be separated into two parts in the weak value expansion: $\gamma_n(C) = \oint \mathbf{A}_n^S \cdot d\mathbf{R} + \oint \mathbf{A}_n^{MP} \cdot d\mathbf{R}$. We emphasize that this decomposition itself is not arbitrary but determined by the pre-defined post-selection state $\langle\varphi(\mathbf{R})|$, while the state $\langle\varphi(\mathbf{R})|$ can be arbitrary selected. For example, the post-selection $\langle\varphi| = \langle u_n|$ vanishes $\mathbf{A}_n^{MP}$, leading to $\mathbf{A}_n = \mathbf{A}_n^S$. The quantity $\mathbf{A}_n^{MP}$ is thus the inherent result of the weak value picture, originating from $\langle\varphi| \neq \langle u_n|$.

With the $\mathbf{R}$-space momentum operator $\mathbf{p_R}$, the vector potential $\mathbf{A}_n^S = -(1/\hbar)[\langle\varphi|\mathbf{p_R}|u_n\rangle/\langle\varphi|u_n\rangle]$ corresponds to the "momentum *weak* value" of the initial eigenstate in the $\mathbf{R}$-space, weakly measuring the canonical momentum of $|u_n\rangle$ with an arbitrarily post-selected state $\langle\varphi(\mathbf{R})|$. On the other hand, with the projection operator to the $m^{\text{th}}$ eigenstate $\mathbf{P}_m = |u_m\rangle\langle u_m|$, $\mathbf{A}_n^{MP}$ is the sum of $\mathbf{A}_n^{MP\text{-}m} = (1/\hbar)\langle\varphi|\mathbf{P}_m\mathbf{p_R}|u_n\rangle/\langle\varphi|u_n\rangle$ ($m \neq n$), which corresponds to the "mutually-*projected* momentum *weak* values" of the $m^{\text{th}}$ eigenstate, weakly measuring the $\mathbf{R}$-space momentum lying along $|u_m\rangle$ with a postselected state $\langle\varphi(\mathbf{R})|$. The reformulation $\mathbf{A}_n = \mathbf{A}_n^S + \mathbf{A}_n^{MP} = -(1/\hbar)\langle u_n|\mathbf{p_R}|u_n\rangle$ then shows that the Berry connection $\mathbf{A}_n$, which is the expectation value of the $\mathbf{R}$-space momentum operator, can be decomposed into the sum of the self ($\mathbf{A}_n^S$) and mutually projected ($\mathbf{A}_n^{MP}$) weak value momenta in the $\mathbf{R}$-space.

Furthermore, the weak value vector potential $\mathbf{A}_n^{MP}$ is an intrinsic quantity originating from the weak value representation because $\mathbf{A}_n^{MP} \neq \mathbf{O}$ only when pre- and post-selected states are different, as $|\varphi\rangle \neq |u_n\rangle$.

With the condition of $\mathbf{A}_n = \mathbf{A}_n^S + \mathbf{A}_n^{MP}$, we further investigate the gauge condition of Eqs. (5) and (6) to answer our original question, "Does each "partial" weak value have the same gauge condition as that of the original expectation value?" First, it is well known [7,8,23] that $\mathbf{A}_n$ is a gauge-dependent quantity, as shown in the change of $\mathbf{A}_n \to \mathbf{A}_n - \nabla_\mathbf{R} \xi_n(\mathbf{R})$ with the gauge transformation of $|u_n\rangle \to exp[i\xi_n(\mathbf{R})]|u_n\rangle$. For the weak value form $\mathbf{A}_n = \mathbf{A}_n^S + \mathbf{A}_n^{MP}$, if we apply the gauge transformation as $|\varphi\rangle \to exp[i\xi_\varphi(\mathbf{R})]|\varphi\rangle$ and $|u_m\rangle \to exp[i\xi_m(\mathbf{R})]|u_m\rangle$, a clear distinction is observed between each weak value vector potential $\mathbf{A}_n^S$ and $\mathbf{A}_n^{MP}$. $\mathbf{A}_n^S$ reflects the apparent gauge dependency as $\mathbf{A}_n^S \to \mathbf{A}_n^S - \nabla_\mathbf{R} \xi_n(\mathbf{R})$, which is the same as that in the original Berry connection $\mathbf{A}_n \to \mathbf{A}_n - \nabla_\mathbf{R} \xi_n(\mathbf{R})$. In contrast, $\mathbf{A}_n^{MP}$ and its elements $\mathbf{A}_n^{MP-m}$ represent the full gauge invariance without any ambiguity, as $\mathbf{A}_n^{MP} \to \mathbf{A}_n^{MP}$ and $\mathbf{A}_n^{MP-m} \to \mathbf{A}_n^{MP-m}$ (see Appendix C for the detailed proof).

This gauge transformation shows that the gauge dependence of the Berry connection $\mathbf{A}_n$ originates exclusively from $\mathbf{A}_n^S$. The gauge invariance *only* for well-defined modulo $2\pi$ of the Berry phase $\oint \mathbf{A}_n d\mathbf{R}$ then originates from the gauge condition of the closed integral of $\mathbf{A}_n^S$ as $\oint \mathbf{A}_n^S \cdot d\mathbf{R}$. In contrast, $\mathbf{A}_n^{MP}$, and more generally, *any* combined operation involving elemental weak values $\mathbf{A}_n^{MP-m}$, constitute locally measurable quantities due to its full gauge invariance. This finding demonstrates that when the nonzero condition of $\mathbf{A}_n^{MP}$ ($|\varphi\rangle \neq |u_n\rangle$) is satisfied, we can extract new measurable (or gauge-invariant) quantities from the originally gauge-dependent one, by exploiting the nonconservation of the gauge condition during the weak-value decomposition. In this context, we can also expect the generalization of this "weak value picture" to the extraction of measurable quantities from other gauge-dependent ones.

## 3. Gauge-dependent weak value for Berry phase

To illustrate the impact of separating gauge-dependent and gauge-invariant parts in the Berry connection $\mathbf{A}_n$, as an example, we restrict our discussion to the post-selection by the "parameter state" $\langle\varphi(\mathbf{R})| = \langle\mathbf{R}|$. For such a particular type of the post-selection, the eigenstate $|u_n(\mathbf{R})\rangle$ (or its linear superposition $|\psi(\mathbf{R})\rangle$) and the parameter state $|\mathbf{R}\rangle$ have to be included in the same Hilbert space, satisfying the relation of $\langle\mathbf{R}|u_n(\mathbf{R})\rangle=u_n(\mathbf{R})$: the R-space wavefunction representation of the state $|u_n(\mathbf{R})\rangle$.

The most illustrative example of the parameter state for the post-selection can be found in the Hilbert space of a spinless particle, where the inner product between a position basis $|\mathbf{r}\rangle$ and a wavefunction $|\psi\rangle$ is defined. Because the Hamiltonian $H$ of a spinless particle is described by the position representation $H = H(\mathbf{r})$, we can set the parameter state as $|\mathbf{R}\rangle = |\mathbf{r}\rangle$, and a scalar wavefunction is defined by the inner product $\langle\mathbf{r}|\psi\rangle = \psi(\mathbf{r})$. The time-dependent variation of the parameter state $|\mathbf{R}(t)\rangle = |\mathbf{r}(t)\rangle$ then represents the time-dependent coordinate transformation, which describes the moving frame or moving potential, which will be shown in the example of Fig. 1.

The application of a position basis as the post-selection $\langle\varphi(\mathbf{R})| = \langle\mathbf{R}| = \langle\mathbf{r}|$ has been widely studied in the subfield of weak value theory, such as superoscillation [33-36] that is described well in Berry's weak value definition of local momentum [37]. Furthermore, as shown in the definition of a wavefunction in reciprocal space $\psi(\mathbf{p}) = \langle\mathbf{p}|\psi\rangle$, the parameter state for a lattice momentum $\langle\mathbf{R}| = \langle\mathbf{p}|$ can also be defined in the momentum representation. To describe this generality, in the later discussion, we maintain the notation of $\langle\mathbf{R}|$ as the parameter state for the post-selection.

Equations (5) and (6) with the post-selection of $\langle\varphi(\mathbf{R})| = \langle\mathbf{R}|$ result in the forms of

$$\mathbf{A}_n^S = i \frac{\nabla_\mathbf{R} u_n(\mathbf{R})}{u_n(\mathbf{R})}, \tag{7}$$

$$\mathbf{A}_n^{MP} = -i \sum_{m \neq n} \frac{u_m(\mathbf{R}) \langle u_m | \nabla_\mathbf{R} | u_n \rangle}{u_n(\mathbf{R})}, \tag{8}$$

From Eqs. (7,8), we now demonstrate that each weak value of $\mathbf{A}_n^S$ and $\mathbf{A}_n^{MP}$ plays a distinct role in the emergence of quantum geometric quantities: Berry phase and curvature. Although Berry connection and Berry curvature are interrelated through $\mathbf{B}_n = \nabla_\mathbf{R} \times \mathbf{A}_n$, it does not mean that the nonzero $\mathbf{B}_n$ should be applied to a particle directly for the emergence of Berry phase obtained from Berry connection as $\gamma_n(C) = \oint \mathbf{A}_n d\mathbf{R}$, as demonstrated in the singular existence of the magnetic field in the Aharonov-Bohm effect [38] and recent work in the momentum space [39]. This condition implies the possibility of controlling each geometric quantity (Berry phase and curvature) independently in a given regime, by separating the source of each geometric quantity in terms of the decomposed parts of the Berry connection $\mathbf{A}_n^S$ and $\mathbf{A}_n^{MP}$.

For this purpose, we revisit a representative example of the Berry phase: the Aharonov-Bohm effect [38] that connects electromagnetic and geometric vector potentials. Consider a solenoid with a uniform magnetic field $\mathbf{B}_0$ (Fig. 1) and $\mathbf{B} = 0$ outside the solenoid, having the singular spatial distribution. For the evolution of a charged particle, the arbitrary scalar potential $V(\mathbf{r} - \mathbf{R}(t))$ that confines the particle [7,23] is assumed for the coordinate origin $\mathbf{R}(t)$. The Hamiltonian $H$, including the nonzero *electromagnetic* vector potential $\mathbf{A}$, is then

$$H = \frac{1}{2m}\left[\frac{\hbar}{i}\nabla_\mathbf{r} - q\mathbf{A}(\mathbf{r})\right]^2 + V(\mathbf{r} - \mathbf{R}), \qquad (9)$$

for the eigenvalue equation $H|u_n(\mathbf{r},\mathbf{R})\rangle = E_n|u_n(\mathbf{r},\mathbf{R})\rangle$ with $\mathbf{R}(t)$, which is the parameter space of the Berry phase in this example defined by the position of a particle. The post-selection of $\langle\varphi(\mathbf{R})| = \langle\mathbf{R}|$ in this example then corresponds to the weak measurement of a certain quantum state along the time-varying position of the charged particle $\mathbf{R}(t)$.

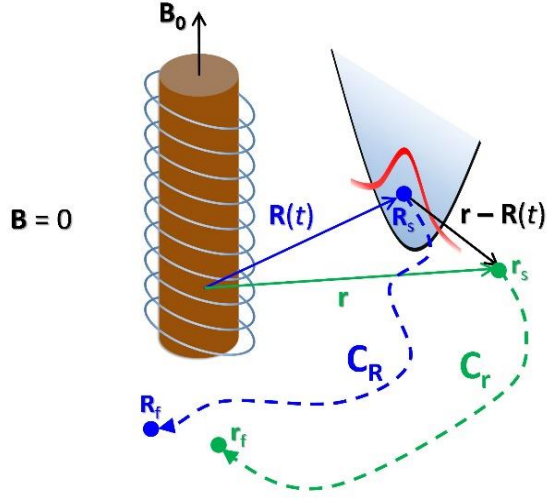

**Figure 1.** Illustration of the Aharonov-Bohm effect to describe the gauge invariance of the vector potential $\mathbf{A}_n^{MP}$. $\mathbf{r}$ is the position vector, and $\mathbf{R}(t)$ is the time-dependent evolution of the scalar potential center for $V(\mathbf{r} - \mathbf{R}(t))$ along the contour $C_R$. $C_r$ is the contour for the integral of $\mathbf{A}_n^{MP}$ in real space. The results of the contour integrals along $C_R$ and $C_r$ are shown in Appendix E.

With $\mathbf{B} = 0$, Eq. (9) can be solved by Peierls substitution [23,38] $|u_n(\mathbf{r},\mathbf{R})\rangle = exp[ig(\mathbf{r},\mathbf{R})]|v_n\rangle$, where $g(\mathbf{r},\mathbf{R}) = (q/\hbar)\int_\mathbf{R}^\mathbf{r} \mathbf{A}(\mathbf{r}')d\mathbf{r}'$, leading to the time-independent Schrödinger equation $\{-[\hbar^2/(2m)]\nabla_\mathbf{r}^2 + V(\mathbf{r} - \mathbf{R})\}|v_n\rangle = E_n|v_n\rangle$. It is worth noting that the eigenstate $|v_n\rangle$ is a function of $(\mathbf{r} - \mathbf{R})$, and its complex phase with respect to $\mathbf{R}$ is identical up to constant. With the post-selection of $\langle\varphi(\mathbf{R})| = \langle\mathbf{R}|$, the gauge-invariant vector potential $\mathbf{A}_n^{MP}$ then becomes (see Appendix D for the detailed derivation)

$$\mathbf{A}_n^{MP} = -i\frac{\nabla_\mathbf{R} v_n(\mathbf{r}-\mathbf{R})}{v_n(\mathbf{r}-\mathbf{R})} = -i\nabla_\mathbf{R}\left(\log v_n(\mathbf{r}-\mathbf{R})\right), \qquad (10)$$

while $\mathbf{A}_n^S = (q/\hbar)\mathbf{A} - \mathbf{A}_n^{MP}$, confirming the well-known relation between the geometric ($\mathbf{A}_n^S + \mathbf{A}_n^{MP}$) and electromagnetic ($\mathbf{A}$) vector potentials.

Because the parameter space $\mathbf{R}(t)$ is the *particle position* for the coordinate origin of $V(\mathbf{r} - \mathbf{R}(t))$ with $\nabla_\mathbf{R}|v_n(\mathbf{r} - \mathbf{R})\rangle = -\nabla_\mathbf{r}|v_n(\mathbf{r} - \mathbf{R})\rangle$, $\mathbf{A}_n^{MP}$ corresponds to the measurable complex-valued local momentum of a stationary wavefunction $|v_n\rangle$ following Berry's definition [37], as $\mathbf{A}_n^{MP} = (1/\hbar)\mathbf{p}_\mathbf{R}v_n/v_n = -(1/\hbar)\mathbf{p}_\mathbf{r}v_n/v_n$, where $\mathbf{p}_\mathbf{R} = -i\hbar\nabla_\mathbf{R}$ and $\mathbf{p}_\mathbf{r} = -i\hbar\nabla_\mathbf{r}$. Figure 2 represents examples of complex-valued $\mathbf{A}_n^{MP}$, which are invariant under gauge transformation. The system is composed of an ideal square potential well defined by $V(\mathbf{r} - \mathbf{R})$ around the confined magnetic field $\mathbf{B}_0$ (Fig. 2a). For the bound states $|v_n\rangle$ (Fig. 2b,c), the transverse ($x$-axis) component of $\mathbf{A}_n^{MP}$ becomes purely imaginary (Fig. 2d,e), while it has real solutions for unbound states. Each node of the bound state results in the divergence of $\mathbf{A}_n^{MP}$, which is the imaginary-valued counterpart of the superoscillation [36]: superamplification or superdecay along the transverse plane. The longitudinal ($y$-axis) component of $\mathbf{A}_n^{MP}$ could be real or imaginary, dependent on the eigenvalue $E_n$.

It is also noted that $\mathbf{A}_n^{MP}$ does not affect the accumulation of the Berry phase due to its local nature (see Appendix E); any line integrals of $\mathbf{A}_n^{MP}$ are independent of the speed or trajectory of system evolution ($C_R$ in Fig. 1) or the integral path in real space ($C_r$ in Fig. 1), and thus, the closed loop leads to zero for these integrals independent from the presence of the electromagnetic vector potential $\mathbf{A}$. Therefore, the Berry phase, which is purely geometric along the closed loop, is solely determined by the unobservable quantity $\mathbf{A}_n^S$, the gauge-dependent part in the Berry connection $\mathbf{A}_n$, as $\gamma_n(C) = \oint \mathbf{A}_n \cdot d\mathbf{R} = \oint \mathbf{A}_n^S \cdot d\mathbf{R}$.

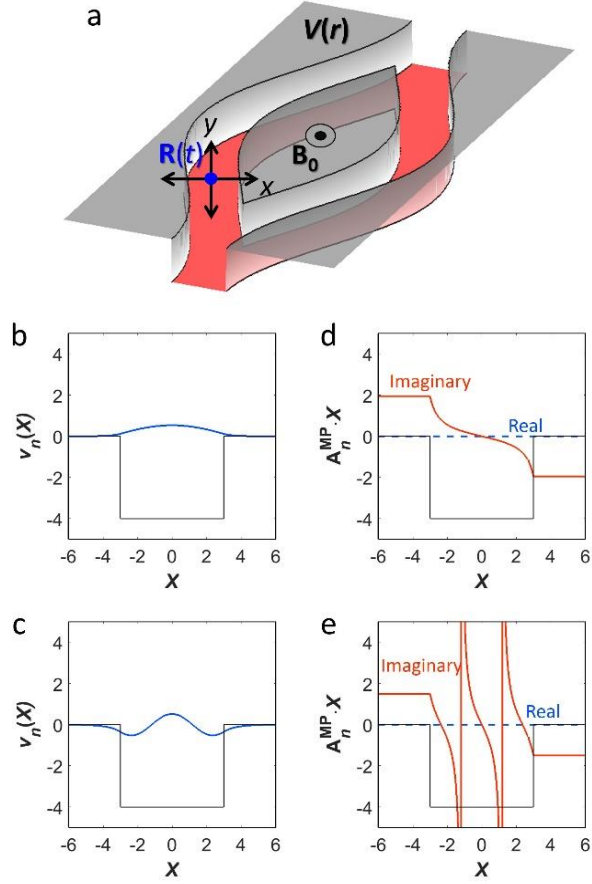

**Figure 2.** Examples of the gauge-invariant vector potential $\mathbf{A}_n^{MP}$. (a) A schematic for the Aharonov-Bohm effect with an ideal square potential well defined by $V(\mathbf{r} - \mathbf{R})$. (b,c) Bound states $|v_n\rangle$ and (d,e) the corresponding $x$-axis component of $\mathbf{A}_n^{MP}$ which is purely imaginary: (b,d) 0$^{th}$ mode and (c,e) 2$^{nd}$ mode. Blue dashed (or orange solid) lines denote real (or imaginary) parts in (d,e). Black solid lines in (b-e) present the shape of the potential. The $x$-axis (or $y$-axis) denotes the transverse (or longitudinal) direction to the evolution of $\mathbf{R}(t)$. The spatial coordinate is normalized as X = $(2m/\hbar)x$. The $y$-axis component of $\mathbf{A}_n^{MP}$ is determined by the eigenvalue.

## 4. Gauge-invariant weak value for Berry curvature

In contrast to the dominant role of $\mathbf{A}_n^S$ in the Berry phase, the role of $\mathbf{A}_n^{MP}$ from the post-selection by $\langle\mathbf{R}|$ is confirmed with the Berry curvature. For the nonzero Berry curvature $\mathbf{B}_n = \nabla_\mathbf{R} \times \mathbf{A}_n$ (e.g., nonzero $\mathbf{B}$ outside the solenoid in Fig. 1), we cannot utilize Peierls substitution $|u_n(\mathbf{r},\mathbf{R})\rangle = exp[ig(\mathbf{r},\mathbf{R})]|v_n\rangle$ for the zero magnetic field because the line integral of $g(\mathbf{r},\mathbf{R}) = (q/\hbar)\int_\mathbf{R}^\mathbf{r} \mathbf{A}(\mathbf{r}')d\mathbf{r}'$ depends on the path [23]. Instead, from Eqs. (7) and (8), the weak value representation of the Berry curvature $\mathbf{B}_n = \nabla_\mathbf{R} \times \mathbf{A}_n$ is decomposed into two parts as $\mathbf{B}_n = \mathbf{B}_n^S + \mathbf{B}_n^{MP}$, where $\mathbf{B}_n^S = \nabla_\mathbf{R} \times \mathbf{A}_n^S$ and $\mathbf{B}_n^{MP} = \nabla_\mathbf{R} \times \mathbf{A}_n^{MP}$. From the relation of $\mathbf{B}_n^S = \mathbf{O}$, it is noted that the Berry curvature exclusively originates from the gauge-invariant vector potential as $\mathbf{B}_n = \mathbf{B}_n^{MP} = \nabla_\mathbf{R} \times \mathbf{A}_n^{MP}$ (see Appendix F for the condition of the zero curvature of $\mathbf{A}_n^S$ for the general post-selection), where the partial weak value for the $m^{th}$ eigenstate ($m \neq n$) leads to the partial Berry curvature $\mathbf{B}_n^{MP-m}$:

$$\mathbf{B}_n^{MP\text{-}m} = -i\frac{u_m(\mathbf{R})}{u_n(\mathbf{R})} \times \left[ \langle\nabla_\mathbf{R} u_m| \times |\nabla_\mathbf{R} u_n\rangle - \nabla_\mathbf{R}\left(\log\frac{u_m(\mathbf{R})}{u_n(\mathbf{R})}\right) \times \langle u_m|\nabla_\mathbf{R}|u_n\rangle \right], \qquad (11)$$

and $\mathbf{B}_n$ is the sum of $\mathbf{B}_n^{MP\text{-}m}$. This is also an alternative expression of the Berry curvature [8] originating from the weak value formulation using the post-selection of $\langle\varphi(\mathbf{R})| = \langle\mathbf{R}|$. We emphasize that the exclusive role of the measurable vector potential $\mathbf{A}_n^{MP}$ in determining the Berry curvature $\mathbf{B}_n$, and the exclusive role of the unobservable vector potential $\mathbf{A}_n^S$ in determining the Berry phase $\gamma_n$ do not contradict with the relationship $\mathbf{B}_n = \nabla_\mathbf{R} \times \mathbf{A}_n$ or its Stokes integral form. Because the nonzero Berry curvature is not a necessary condition for the Berry phase [39] as shown in an Aharonov-Bohm [38] or molecular Aharnovo-Bohm [40] effects, for example, zero Berry curvature (with $\mathbf{A}_n^{MP} = \mathbf{O}$) can also provide the Berry phase (with $\mathbf{A}_n^S \neq \mathbf{O}$).

## 5. Conclusion

In conclusion, we have shown that the definition of the gauge-invariant vector potential $\mathbf{A}_n^{MP}$ can be extracted from the weak value decomposition of the gauge-dependent Berry connection $\mathbf{A}_n$ in terms of the momentum representation in the complex-Hilbert parameter space. With the post-selection of the parameter space state $\langle \mathbf{R}|$, we demonstrated the local nature of $\mathbf{A}_n^{MP}$ independent of the "speed" or "trajectory" of system evolution in the Aharonov-Bohm effect, which exhibits the exclusive vector potential origin $\mathbf{A}_n^S$ of the Berry phase. We also proved that $\mathbf{A}_n^{MP}$ plays an exclusive role in the emergence of the Berry curvature, as $\mathbf{B}_n = \nabla \times \mathbf{A}_n^{MP}$.

In the context of finding a measurable quantity from an unobservable one, the suggested decomposition process using the concept of weak values is distinct from closed-loop integrals for geometric phases [7,12,14] or differential operators applied to scalar and vector potentials in electromagnetics [1,2]. The separation of the gauge-dependent and gauge-invariant parts of the Berry connection suggests that this process will enable the decomposition of a general gauge-dependent quantity into partial weak values, which may include gauge-invariant, measurable quantities. Considering the significant roles of the Berry connection, phase and curvature in topological phenomena, a new viewpoint and representation of the vector potential origins of the Berry phase and curvature will provide a new framework for the topological understanding of quantum mechanics [8,10] and optics [9,11] in terms of the independent control of quantum geometric quantities. Due to the emergence of the momentum weak value, which has played a critical role in the de Broglie-Bohm representation [41,42] and Madelung's hydrodynamic form of quantum mechanics [43], this result will also stimulate a geometric view on Bohmian mechanics and hydrodynamic representation. Furthermore, applying the weak-value decomposition to an arbitrary quantum-mechanical operator can also be envisaged, not only to reveal a new gauge-invariant quantity but also to subdivide existing quantities with new physical pictures, as shown in $\mathbf{A}_n^{MP}$ and its role in the decomposition of quantum geometric quantities.

## Acknowledgements

This work was supported by the New Faculty Startup Fund from Seoul National University and the National Research Foundation of Korea (NRF) through the Global Frontier Program (GFP, 2014M3A6B3063708) funded by the Korean government. S. Yu was also supported by the National Research Foundation of Korea (NRF) grant funded by the Korea government(MSIT) (No. 2021R1C1C1005031).

## Appendix A. Hermitian condition of R-space momentum operator

As the generalization of the relation between the spatial translation operator and canonical momentum operator [24], the $\mathbf{R}$-translation operator $\mathbf{T_R}(\Delta R_m)$ for the $m^{\text{th}}$ parameter $R_m$ can be expressed as

$$\mathbf{T}(\Delta R_m) = 1 - \frac{i}{\hbar} p_{\mathbf{R}}^m \Delta R_m + \cdots. \tag{S1}$$

Equation (S1) becomes Eq. (3) in the main text, or $\mathbf{T_R}(\Delta R_m) = 1 - (i/\hbar)p_{\mathbf{R}}^m \Delta R_m$, for the infinitesimal translation $\Delta R_m$. This relation leads to $\mathbf{T_R}(\Delta R_m)^\dagger = 1 + (i/\hbar)p_{\mathbf{R}}^{m\dagger}\Delta R_m$, while $\mathbf{T_R}(-\Delta R_m) = \mathbf{T_R}(\Delta R_m)^{-1} = 1 + (i/\hbar)p_{\mathbf{R}}^m \Delta R_m$. Therefore, the Hermitian condition of the $\mathbf{R}$-space momentum operator $p_{\mathbf{R}}^{m\dagger} = p_{\mathbf{R}}^m$ is achieved with the unitary operator $\mathbf{T_R}(\Delta R_m)$, as $\mathbf{T_R}(\Delta R_m)^\dagger = \mathbf{T_R}(\Delta R_m)^{-1} = \mathbf{T_R}(-\Delta R_m)$. The application of the weak value theory [16-18,32], which requires the Hermitian condition of the operator, is then allowed with the unitary translation of the wavefunction on the parameter space $\mathbf{R}$.

## Appendix B. Direct application of $\langle \varphi(\mathbf{R})|$ to the time-dependent Schrödinger equation

The result of Eq. (4) in the main text can also be derived by applying the post-selection with $\langle \varphi(\mathbf{R})|$ in place of $\langle u_n(\mathbf{R})|$ to the state of the time-dependent Schrödinger equation. From Eq. (2) in the main text, we obtain

$$\langle \varphi | \frac{\partial}{\partial t} | u_n \rangle + i\frac{\partial \gamma_n}{\partial t}\langle \varphi | u_n \rangle = -e^{-i(\theta_n + \gamma_n)}\varepsilon \sum_{m \neq n}\left(\frac{i}{\hbar}c_m E_m + \frac{\partial c_m}{\partial t}\right)\langle \varphi | u_m \rangle. \tag{S2}$$

The coefficients on the right side of Eq. (S2) are obtained by multiplying Eq. (2) in the main text by the bra eigenstate $\langle u_p(\mathbf{R})|$ ($p \neq n$) as follows:

$$\langle u_p | \frac{\partial}{\partial t} | u_n \rangle = -e^{-i(\theta_n + \gamma_n)}\varepsilon\left(\frac{i}{\hbar}c_p E_p + \frac{\partial c_p}{\partial t}\right). \tag{S3}$$

Using Eq. (S3), we can re-express Eq. (S2) as follows:

$$\langle\varphi|\frac{\partial}{\partial t}|u_n\rangle + i\frac{\partial\gamma_n}{\partial t}\langle\varphi|u_n\rangle = \sum_{m\neq n}\langle u_m|\frac{\partial}{\partial t}|u_n\rangle\langle\varphi|u_m\rangle = \sum_{m\neq n}\langle\varphi|u_m\rangle\langle u_m|\frac{\partial}{\partial t}|u_n\rangle. \tag{S4}$$

The time derivative of the Berry phase $\partial_t\gamma_n$ then takes the form of Eq. (4) in the main text with the condition of $\langle\varphi|u_n\rangle \neq 0$.

## Appendix C. Gauge conditions of the weak value vector potentials

To investigate the gauge condition of each weak value vector potential, we apply the gauge transformation to the weak value form $\mathbf{A}_n = \mathbf{A}_n^S + \mathbf{A}_n^{MP}$, as $|\varphi\rangle \to exp[i\xi_\varphi(\mathbf{R})]|\varphi\rangle$, $|u_m\rangle \to exp[i\xi_m(\mathbf{R})]|u_m\rangle$, and $|u_n\rangle \to exp[i\xi_n(\mathbf{R})]|u_n\rangle$. The momentum weak value $\mathbf{A}_n^S$ then becomes

$$\mathbf{A}_n^{\text{S-gauge}} = i\frac{\langle\varphi|e^{-i\xi_\varphi(\mathbf{R})}\nabla_\mathbf{R} e^{i\xi_n(\mathbf{R})}|u_n\rangle}{\langle\varphi|e^{-i\xi_\varphi(\mathbf{R})}e^{i\xi_n(\mathbf{R})}|u_n\rangle} = i\frac{e^{i[\xi_n(\mathbf{R})-\xi_\varphi(\mathbf{R})]}\langle\varphi|(\nabla_\mathbf{R}|u_n\rangle + i|u_n\rangle\nabla_\mathbf{R}\xi_n(\mathbf{R}))}{e^{i[\xi_n(\mathbf{R})-\xi_\varphi(\mathbf{R})]}\langle\varphi|u_n\rangle} = \mathbf{A}_n^S - \nabla_\mathbf{R}\xi_n(\mathbf{R}), \tag{S5}$$

which illustrates the identical gauge dependency $-\nabla_\mathbf{R}\xi_n(\mathbf{R})$ to that of the original Berry connection $\mathbf{A}_n$. In contrast, the projected momentum weak value of the $m^{\text{th}}$ eigenstate $\mathbf{A}_n^{MP\text{-}m}$ becomes

$$\begin{aligned}\mathbf{A}_n^{MP\text{-}m\text{-gauge}} &= -i\frac{\langle\varphi|e^{-i\xi_\varphi(\mathbf{R})}e^{i\xi_m(\mathbf{R})}|u_m\rangle\langle u_m|e^{-i\xi_m(\mathbf{R})}\nabla_\mathbf{R} e^{i\xi_n(\mathbf{R})}|u_n\rangle}{\langle\varphi|e^{-i\xi_\varphi(\mathbf{R})}e^{i\xi_n(\mathbf{R})}|u_n\rangle} \\ &= -i\frac{e^{i[\xi_n(\mathbf{R})-\xi_\varphi(\mathbf{R})]}\langle\varphi|u_m\rangle\langle u_m|(\nabla_\mathbf{R}|u_n\rangle + i|u_n\rangle\nabla_\mathbf{R}\xi_n(\mathbf{R}))}{e^{i[\xi_n(\mathbf{R})-\xi_\varphi(\mathbf{R})]}\langle\varphi|u_n\rangle} \\ &= -i\frac{\langle\varphi|u_m\rangle\langle u_m|\nabla_\mathbf{R}|u_n\rangle + i\langle u_m|u_n\rangle\nabla_\mathbf{R}\xi_n(\mathbf{R})}{\langle\varphi|u_n\rangle} = \mathbf{A}_n^{MP\text{-}m}.\end{aligned} \tag{S6}$$

Because the $\mathbf{A}_n^{MP\text{-}m}$ values for all $m \neq n$ are gauge-invariant, their sum $\mathbf{A}_n^{MP}$ is also gauge-invariant.

## Appendix D. Derivation of $\mathbf{A}_n^{MP}$ for the Aharonov-Bohm effect

We start from $|u_{n,m}(\mathbf{r},\mathbf{R})\rangle = exp[ig(\mathbf{r},\mathbf{R})]|v_{n,m}\rangle$, where $g(\mathbf{r},\mathbf{R}) = (q/\hbar)\int_\mathbf{R}^\mathbf{r}\mathbf{A}(\mathbf{r}')d\mathbf{r}'$ is defined by the electromagnetic vector potential $\mathbf{A}$ and $\nabla_\mathbf{R} g(\mathbf{r},\mathbf{R}) = -(q/\hbar)\mathbf{A}(\mathbf{R})$. $\mathbf{A}_n^{MP\text{-}m}$ then becomes

$$\begin{aligned}\mathbf{A}_n^{MP} &= -i\sum_{m\neq n}\frac{\langle\varphi|e^{ig(\mathbf{r},\mathbf{R})}|v_m\rangle\langle v_m|e^{-ig(\mathbf{r},\mathbf{R})}(\nabla_\mathbf{R}|v_n\rangle + i|v_n\rangle\nabla_\mathbf{R}g(\mathbf{r},\mathbf{R}))e^{ig(\mathbf{r},\mathbf{R})}}{\langle\varphi|e^{ig(\mathbf{r},\mathbf{R})}|v_n\rangle} \\ &= -i\sum_{m\neq n}\frac{\langle\varphi|v_m\rangle\langle v_m|(\nabla_\mathbf{R}|v_n\rangle + i|v_n\rangle\nabla_\mathbf{R}g(\mathbf{r},\mathbf{R}))}{\langle\varphi|v_n\rangle} \\ &= -i\sum_{m\neq n}\frac{\langle\varphi|v_m\rangle\langle v_m|\nabla_\mathbf{R}|v_n\rangle}{\langle\varphi|v_n\rangle} \\ &= -i\sum_m\frac{\langle\varphi|v_m\rangle\langle v_m|\nabla_\mathbf{R}|v_n\rangle}{\langle\varphi|v_n\rangle} + i\langle v_n|\nabla_\mathbf{R}|v_n\rangle \\ &= -i\frac{\langle\varphi|\nabla_\mathbf{R}|v_n\rangle}{\langle\varphi|v_n\rangle} + i\langle v_n|\nabla_\mathbf{R}|v_n\rangle.\end{aligned} \tag{S7}$$

Because $\nabla_\mathbf{R}|v_n(\mathbf{r}-\mathbf{R})\rangle = -\nabla_\mathbf{r}|v_n(\mathbf{r}-\mathbf{R})\rangle$, we can change the second term in Eq. (S7) to the expectation value of the momentum $i\langle v_n|\nabla_\mathbf{R}|v_n\rangle = -i\langle v_n|\nabla_\mathbf{r}|v_n\rangle = \langle v_n|\mathbf{p}_\mathbf{r}|v_n\rangle/\hbar$, which becomes zero [23] in the stationary state $|v_n\rangle$. Therefore, by applying the post-selected state $\langle\varphi(\mathbf{R})| = \langle\mathbf{R}|$, Eq. (S7) becomes

$$\mathbf{A}_n^{MP} = -i\frac{\langle\mathbf{R}|\nabla_\mathbf{R}|v_n\rangle}{\langle\mathbf{R}|v_n\rangle} = -i\frac{\nabla_\mathbf{R} v_n(\mathbf{r}-\mathbf{R})}{v_n(\mathbf{r}-\mathbf{R})}, \tag{S8}$$

leading to Eq. (10) in the main text.

## Appendix E. Contour integral of $\mathbf{A}_n^{MP}$

To examine the effects of $\mathbf{A}_n^{MP}$ for $\gamma_n^{MP} = \int\mathbf{A}_n^{MP}d\mathbf{R}$, we define two gauge-invariant phase evolutions $\gamma_n^{MP\text{-}\mathbf{r}}$ and $\gamma_n^{MP\text{-}\mathbf{R}}$ driven by $\mathbf{A}_n^{MP}$, for arbitrary (including non-cyclic) contours, as follows:

$$\gamma_n^{MP\text{-}\mathbf{r}}(\mathbf{r})\big|_{C_R} = -i\int_{C_R}\nabla_\mathbf{R}(\log v_n(\mathbf{r}-\mathbf{R}))d\mathbf{R} = -i\log\frac{v_n(\mathbf{r}-\mathbf{R}_f)}{v_n(\mathbf{r}-\mathbf{R}_s)}, \tag{S9}$$

$$\gamma_n^{\text{MP-R}}(\mathbf{R})\bigg|_{C_r} = i\int_{C_r} \nabla_\mathbf{r}\left(\log v_n(\mathbf{r}-\mathbf{R})\right)d\mathbf{r} = -i\log\frac{v_n(\mathbf{r}_s-\mathbf{R})}{v_n(\mathbf{r}_f-\mathbf{R})}. \tag{S10}$$

These two complex-valued phase functions represent distinct observation conditions for $\mathbf{A}_n^{\text{MP}}$ in terms of the measurement positions ($\mathbf{r}$ or $\mathbf{R}$) and the evolutions ($C_R$ or $C_r$): $\gamma_n^{\text{MP-r}}(\mathbf{r})$ for the phase accumulation at *real-space* position $\mathbf{r}$ with the *system* evolution along $C_R$, and $\gamma_n^{\text{MP-R}}(\mathbf{R})$ for the phase accumulation measured along the *real-space* contour $C_r$ with the *system* position $\mathbf{R}$. By virtue of the gauge invariance of $\mathbf{A}_n^{\text{MP}}$, both phase accumulations are automatically gauge-invariant regardless of whether the contours defined for the potential ($C_R$) or position ($C_r$) are cyclic. Furthermore, because the line integral is defined only with the initial and final state of $v_n(\mathbf{r}-\mathbf{R})$, both $\gamma_n^{\text{MP-r}}$ and $\gamma_n^{\text{MP-R}}$ are independent of the speed or trajectory of system evolution. $\mathbf{A}_n^{\text{MP}}$ therefore does not contribute to the accumulation of the Berry phase, because $\gamma_n^{\text{MP-r}}(\mathbf{r}) = \gamma_n^{\text{MP-R}}(\mathbf{R}) = 0$ for the closed loops $C_R$ and $C_r$ both with the same initial and final state. We also note that the phase induced by the gauge-invariant $\mathbf{A}_n^{\text{MP}}$ in a non-cyclic evolution process is apparently distinct from the case of a Pancharatnam phase [14], which is dependent on the trajectory composing the closed curve with the geodesic curve in the ray space.

### Appendix F. Zero curvature condition of $\mathbf{A}_n^S$

To separate the source of the Berry curvature in terms of weak-value vector potentials, it is necessary to assign $\mathbf{B}_n^S = \nabla_\mathbf{R} \times \mathbf{A}_n^S = \mathbf{O}$ by introducing suitable post-selections. From Eq. (5) in the main text, the general expression of $\mathbf{B}_n^S$ is

$$\mathbf{B}_n^S = \nabla_\mathbf{R} \times \mathbf{A}_n^S = i\frac{\langle\nabla_\mathbf{R}\varphi|\times|\nabla_\mathbf{R}u_n\rangle - \langle\nabla_\mathbf{R}\varphi|u_n\rangle\times\langle\varphi|\nabla_\mathbf{R}u_n\rangle}{\langle\varphi|u_n\rangle}, \tag{S11}$$

which results in the zero curvature condition of $\langle\nabla_\mathbf{R}\varphi|\times|\nabla_\mathbf{R}u_m\rangle = \langle\nabla_\mathbf{R}\varphi|u_m\rangle\times\langle\varphi|\nabla_\mathbf{R}u_m\rangle$. It is noted that the specific post-selection by the parameter state $\langle\varphi(\mathbf{R})| = \langle\mathbf{R}|$ satisfies this condition.